# An Optically Addressable Transmissive Liquid Crystal Metasurface Spatial Light Modulator

Jared Sisler[1], Claudio U. Hail[1], Zoey S. Davidson[2], Austin M. K. Fehr[2], Jiannan Gao[2], Ruzan Sokhoyan[1], Selim Elhadj[2], and Harry A. Atwater[1]

[1] *Thomas J. Watson Laboratories of Applied Physics, California Institute of Technology, Pasadena, California 91125, USA*

[2] *Seurat Technologies Inc., Wilmington, Massachusetts 01887, USA*

Corresponding authors:
Selim Elhadj and Harry A. Atwater

**Abstract**

Active wavefront control in high-power laser illumination systems is important for technologies such as additive manufacturing, free-space laser communication, and power transmission. Conventional spatial light modulators (SLMs) and mechanical beam-steering devices are unsuitable for such applications as they rely on metal mirrors and electrical contacts which are damaged under high laser irradiances. Here, we report on the design and realization of an optically addressable metasurface liquid crystal (LC)-based SLM for the modulation of high-power transmitted light. Our device uses a photoactive top contact which is optically addressed with a patterned 435 nm laser, creating a transient electrical contact that selectively switches the underlying LC medium. A $TiO_2$ metasurface, resonant in the 915 – 985 nm wavelength range, is embedded within a thin (~2 $\mu$m) LC layer and enables large optical tunability. We demonstrate 90° linear polarization rotation in reconfigurable patterns across a 5x5 $mm^2$ active area with an overall transmittance of > 60%. Additionally, we develop a multiphysics approach to simulate transmittance modulation in our device by modeling the LC interactions with $TiO_2$ nanopillars under an applied electrostatic field. This model exhibits good agreement with measurements and provides improved understanding of how LCs interact with both transmitted light and nanoscale metastructures in active devices. We show that our design and fabrication approach can yield high-performing transmissive metasurface SLM devices and lay the groundwork for the design of future LC-based active nanophotonics.

## 1. Introduction

Dynamic manipulation of the wavefront of light is important to many technologies such as holography [1], beam shaping/steering [2,3], imaging [4], computation [5], and communication [6]. The two most common devices for active wavefront control in two dimensions (2D) are the digital micromirror device (DMD) and liquid crystal (LC) on silicon (LCoS)-based spatial light modulator (SLM). While DMDs have advantages in efficiency and bandwidth, they typically can only achieve binary amplitude modulation, limiting their functionality [7]. LCoS SLMs, however, can achieve > $2\pi$ greyscale phase modulation in the visible and near-infrared (near-IR) wavelength range, making them suitable for more complex optical functions [8,9]. While LCoS SLMs represent the state-of-the-art for active wavefront control, they are reaching fundamental limits of switching speeds and power handling capabilities [10–12]. In this work, we present a metasurface-enhanced LC-based SLM to surpass these limitations.

A conventional LCoS SLM uses a layer of well-aligned LCs sandwiched between a backplane of electrically addressable pixelated contacts and a uniform top transparent contact (indium-tin-oxide) to shape the phase, amplitude, or polarization of reflected light across a 2D plane. Its operation employs the field-controlled orientation and birefringence of LCs to induce a change in accumulated phase of reflected light within a given pixel when the voltage is turned on (liquid crystals typically aligned along electric field vector) or turned off (liquid crystals typically aligned in plane). This functionality allows for >$2\pi$ phase shift with high overall reflectivity and relatively slow switching speeds (~1 kHz) determined by the LC alignment and relaxation dynamics [8].

The switching speed of a single pixel in a LCoS SLM is limited by its relaxation time, $\tau_{OFF} \propto d^2$, where $d$ is the thickness of the LC layer [13]. Thus, by decreasing the thickness of the LC layer, the modulation frequency can be increased. However, because the phase shift from a pixel is non-resonant and relies on propagation phase, there is a minimum thickness of the LC layer required to achieve a desired phase, amplitude, or polarization contrast. This thickness limits the maximum achievable switching speed. Additionally, due to small amounts of absorption in the reflecting back layer, conventional reflective LCoS SLMs are not suitable for use in high-power laser systems. An ideal device for modulation of high-power lasers would be fully transmissive, thus removing any absorption from reflective metallic layers which can cause significant heating and damage. However, in transmissive SLM design, the LC layer must be approximately doubled to accumulate the required phase in a single pass of light. This doubling of LC thickness, $d$, results in a four times increase of switching time, $\tau_{OFF}$. Additionally, creating a transmissive 2D array of electrically addressable contacts presents significant challenges to handle exposure to a high-power beam and often results in larger pixel sizes and decreased overall efficiency [8]. Thus, conventional LCoS SLMs designs are reaching limits in important performance metrics such as speed and resolution for high power handling applications, especially for operation in transmission.

To address current limitations of LCoS SLMs, researchers have immersed an array of nanostructured resonators into a LC layer to locally increase light's interaction with LCs [10,11,14–18]. By doing this, the device's optical response is dominated by the dynamics of LCs in a smaller volume surrounding the resonators, thus allowing the thickness of the LC layer to be decreased, resulting in increased switching speeds. While these devices have shown impressive performance in simulation, there has been a gap between simulation and experiment. This is a common trend among all LC-actuated nanophotonic devices supporting strong optical resonances. It has been proposed that this discrepancy is caused by the complex nanoscale alignment of LCs around nanostructures [18,19]. This is a difficult problem to address as small variations in the LC alignment can result in large changes in optical response due to the metasurface-enhanced optical field. Therefore, there is a need to understand and precisely model the alignment of LCs around nanostructures through thermodynamic free-energy minimization simulations [20–22]. The standard method for modeling the alignment of LCs in complex geometries uses the Landau-de Gennes equations [23]. However, a comprehensive model of the dynamics of LCs on the nanoscale has not yet been achieved in the context of LC-tuned active nanophotonic devices.

While resonantly enhanced LC-based SLMs have enabled faster switching speeds and smaller pixel sizes, most still rely on metallic layers for reflection and electrical contacts [11,16]. Due to the relatively low damage threshold of metals caused by their free-carrier absorption (7.9 J/cm$^2$ for $\lambda$ = 1 $\mu$m illumination in gold [24]), they are not suitable for use with high-power lasers [25]. There are many technologies which require high laser irradiances such as additive manufacturing [26], free-space laser communication [27], and optical power transmission [28]. Additionally, as the cost and size of high-power lasers decrease, their use in other applications becomes more common [29].

In contrast to the power-limited LCoS devices, the optically addressable light valve (OALV) has been previously reported as a LC-based SLM capable of modulating high-power light in 2D [30,31]. This device uses a photo-addressable top contact, effectively an optically addressed capacitive voltage divider, to control LCs for the patterned modulation of transmitted light. While such a device can be used for high-power lasers, it suffers from the same switching speed and pixel size limitations as conventional LCoS SLMs. Thus, we propose the introduction of resonant nanostructures in an OALV architecture to address the outlined limitations of transmissive LC-based SLMs.

Here, we present a millimeter-scale, transmissive, optically addressable SLM using the reorientation of LCs coupled to a titanium dioxide ($TiO_2$) metasurface for control over the polarization of a high-power laser in 2D. By integrating Mie-resonant $TiO_2$ metastructures into an OALV, we can use a thinner LC layer than would typically be required and enable faster switching. In addition, we implement an all-dielectric transparent material stack in the device designed to minimize losses and laser absorption. We provide detailed modeling of the LCs alignment around metasurface structures under an applied electrostatic field which allows us to provide recommendations for future nanophotonic devices utilizing LCs. This work introduces a new platform for all-dielectric, optically addressable transmissive SLMs operating with high-power lasers and furthers the understanding of an emerging field of active nanophotonic devices using LCs.

## 2. Results

**Active device architecture**

A conceptual schematic of our device is presented in **Fig. 1a**. The active layer consists of a $TiO_2$ metasurface submerged in a layer of E7 LCs. The metasurface induces optical resonances which are mostly confined to the $TiO_2$ pillars but slightly extend into the surrounding LC environment. Thus, the modulated optical response is highly sensitive to the tuning and reorientation of LCs within a ~ 200 nm layer around the $TiO_2$ nanostructures. The LC layer is sandwiched between two substrates, each with its own electrical contact. The bottom substrate includes an LC-facing, transparent field-spreading layer of indium tin oxide (ITO). The top substrate includes an optically addressed photoconductor switch consisting of a semi-insulating bismuth silicon oxide (BSO) crystal. BSO has a high electrical impedance in its dark state but generates electrical carriers (decreasing its impedance) when photoexcited with blue light. An electrical bias is constantly applied between the bottom ITO and top BSO substrates. In the dark state, the voltage drop primarily occurs across the semi-insulating BSO layer because of its high impedance. When the BSO is illuminated with patterned blue light from a projector generating an arbitrary pattern, the impedance of the BSO layer is locally decreased in the illuminated regions, selectively causing the voltage to drop across the LC layer and transferring the electric field to the underlying LCs and causing them to rotate. This LC optical addressing scheme forms the basis of the Area Printing [32] for high-power laser 3D metal printing and for beam blocker technology at the National Ignition Facility [33,34]. It allows the 2D patterned switching of a transmissive LC device without any metal interconnects that would be damaged if exposed to a high-power laser beam.

In our experiments, an infrared (IR) linearly polarized, tunable laser ($\lambda \sim$ 915-985 nm) is incident on the device. This IR beam is what we are interested in modulating and is resonant with our designed metasurface. We project an image onto the top BSO layer with blue ($\lambda$ = 435 nm) light, rotating the LCs only in the illuminated regions, effectively transferring the pattern of the low-power blue laser to the polarization of the high-power IR laser. It should be noted that the BSO crystal bandgap (~ 3eV) is larger than the IR light energy (~1.2eV). Thus, the high-power IR illumination does not generate any carriers in the BSO. When the transmitted IR light is then passed through a cross-polarizer at the output, the part of the beam with unaffected polarization is rejected. Thus, our active metasurface SLM has transferred the pattern of the blue pump light to the transmittance of the output IR probe light.

In **Fig. 1a**, the LC layer is presented as a continuous layer which is switched by the blue incident light on the photoconductor. In reality, the LCs have complex orientational local arrangement throughout the cell dictated by local surface interactions, which can be described by a director field in 3D. (LCs experience the same force in their parallel and antiparallel alignment, meaning they can be described by a director field rather than a vector field.) The arrangement of LCs surround nanostructures is not trivial to model and is primarily driven by an anchoring alignment layer on the top substrate bounding the LC layer, as well as by the local wall geometry and surface chemistry of the $TiO_2$ nanostructures. The insets (**Figs. 1b,c**) show simulated LC director fields around the $TiO_2$ nanostructures when the voltage is off and on, respectively. These "$V_{OFF}$" and "$V_{ON}$" states correspond to dark and bright areas of the blue photoexcitation projector/masked beam on the BSO layer, respectively.

The experimental setup used for 2D polarization conversion is shown in **Fig. 1d**. The top beam path shows the blue "write" laser which images a pattern, specified by a physical mask, on the top BSO layer. In our experiments, the mask was a static chrome shadow-mask but could be replaced with an active device such as a reflective SLM or a DMD projector. The bottom light path in **Fig. 1d** shows the IR "probe" laser to which the blue "write" pattern is transferred via polarization modulation, after passing through the device. The IR beam is linearly polarized and first passes through two lenses to expand the beam. All the output light of the metasurface is collected with an objective which then passes through a low-pass filter to remove residual blue light (although most of the pump light is absorbed by the BSO layer), then through a cross-polarizer to only collect the polarization-converted light. This light is then imaged onto a camera, allowing us to view the final polarization-patterned beam.

**Optical characterization of a $TiO_2$ metasurface submerged in liquid crystals**

The cross-section of our device unit cell is shown in **Fig. 2a.** Throughout this work, the "Top Substrate" consists of either ITO and sapphire or a single layer of BSO. To first characterize our devices, we use an ITO and sapphire top substrate in place of the BSO to decouple any effects from the BSO layer, allowing us to characterize the LC response when modulated by a uniform, unpatterned electric field. For demonstration of the optically addressable IR beam control, we built devices using BSO as the top substrate to allow polarization patterning in 2D. An optical image of a final device in our optical setup is shown in **Fig. 2b**. The smaller yellow substrate is the top BSO layer, and the larger transparent, beveled substrate is the back ITO/sapphire. The 5x5 mm$^2$ metasurface is at the center of the device and is shown with three horizontal blue lines projected on the BSO layer to generate a three-line patterning of the co-incident IR beam. Tilted scanning electron micrographs of the $TiO_2$ metasurface are provided in **Figs. 2c-d**. The near vertical sidewalls are obtained from an optimized $SF_6/C_4F_8/Ar$ dry etching process. (See Methods for full fabrication process.)

We first characterized the passive response of our $TiO_2$ metasurface without any LCs. We designed metasurfaces which demonstrate electric and magnetic dipole resonances within the available bandwidth of our Toptica continuously tunable laser ($\lambda$ = 915–985 nm) when embedded in a medium with refractive index approximately equal to the ordinary refractive index of E7 LCs ($n_o$ ~ 1.51 at $\lambda$ = 1 $\mu$m). This surrounding medium was used to provide a better match between the designed modes of our passive metasurface and those that would be excited with embedded LCs. The simulated transmittance spectra of 21 separate metasurfaces of varying nanostructure cylinder diameters is plotted in **Fig. 2e**. All simulated transmittance was obtained using a finite-difference time-domain (FDTD) solver (Lumerical, Inc.). We then fabricated these 21 $TiO_2$ metasurfaces and used poly-methylmethacrylate (PMMA) as the metasurface encapsulating material, which has a refractive index of ~ 1.5 at $\lambda$ = 1 $\mu$m. The measured transmittance spectra of the 21 metasurfaces exposed to the IR beam are plotted in **Fig. 2f**. For all simulation and experiment results, the incident light was linearly polarized in the plane of the $TiO_2$ pillar array and parallel to the alignment direction of the top alignment layer (y-axis in **Figs. 1** and **2**). Each metasurface consists of an array (200x200 $\mu m^2$ in experiment, infinite in simulation) of 310 nm tall $TiO_2$ circular cylinders of varying diameter in a rectangular 2D array with periodicity of 574 nm in each axis. In simulation, we assume an IR incidence angle of 1° off-normal which better represented our measured spectra. The measured and simulated dispersion curves for our passive $TiO_2$ metasurface surrounded by PMMA (**Figs. 2e,f**) show excellent agreement and exhibit four clear resonances in the bandwidth of interest. The optical electric and magnetic field profiles for each simulated mode are shown in Supplementary Information Fig. S1.

Having confirmed our baseline model of the $TiO_2$ metasurfaces was performing as expected when embedded in PMMA, we removed the PMMA, applied a hexamethyldisilazane (HMDS) surface treatment to the $TiO_2$ metasurface, and filled the device with LCs (see Methods). This treatment coats the $TiO_2$ nanostructures and underlying $SiO_2$ with a monolayer of HMDS and has been shown to induce homeotropic (surface normal) alignment of E7 LCs [35]. As we do not use a dedicated alignment layer on the bottom surface of our LC cell, this surface coating helps induce stronger alignment of the LCs. The measured transmittance for all 21 metasurfaces embedded in the LCs is shown in **Fig. 2g** for no applied bias (0 V) and **Fig. 2h** for the maximum applied bias in experiment: a 20 $V_{p-p}$ 1 kHz AC square wave (10 $V_{rms}$). It should be noted that LC devices must always be driven with an AC voltage signal to prevent degradation caused by ion migration in the LCs. As we are driving with a square wave symmetric around 0 V, however, the LCs only respond to the magnitude of the applied signal, not the change in polarity. The maximum applied bias of 10 $V_{rms}$ was selected as our saturation voltage which was experimentally observed (see Supplementary Information Fig. S2).

Next, we present two simulation results corresponding to two different assumptions about how the LCs align around the metasurface structures. The first, labeled "Uniform LC" assumes that at 0 V (**Fig. 2i**), all LCs point along the induced alignment direction (y-axis), and at 10 $V_{rms}$ (**Fig. 2j**), all LCs align perfectly along the direction of the applied electric field (z-axis), perpendicular to the substrate surface. The second LC alignment assumption (**Figs. 2k,l**) uses the Landau-de Gennes thermodynamic equations to calculate how the LCs orient around the $TiO_2$ nanostructures. Additionally, for the 10 $V_{rms}$ case in **Fig. 2l**, we simulate the spatially varying electrostatic field throughout the cell and import this into our Landau-de Gennes simulations to more accurately represent how the LCs switch and reorient under an applied field. Once we obtain a 3D director field describing the LC orientation in space for each voltage, we convert this to the corresponding spatially varying diagonal anisotropic refractive index, which is then imported into Lumerical to simulate the final optical transmittance.

From the data presented in **Fig. 2**, we first note that the measured transmittance with LCs and no applied field (**Fig. 2g**) is significantly different from the measured transmittance with PMMA (**Fig. 2f**). Thus, by changing the surrounding medium of the $TiO_2$ structures from PMMA to LCs, the optical response is changed. By comparing the measured and simulated transmittance at 0 V (**Figs. 2g,i,k**), we see that the Landau-de Gennes simulations give a much closer representation of the measured resonances than the uniform LC field simulations. When the saturation voltage is applied in experiment (**Fig. 2h**), the shortest-wavelength (bluest) mode is modulated while the two longest-wavelength (reddest) modes are not significantly changed from the 0 V case. In simulation at 10 $V_{rms}$, both LC field assumptions begin to converge, producing three similar modes. However, the Landau-de Gennes model predicts a mode which is not captured in the "Uniform LC" assumption, yet is apparent in experiment. Additionally, this mode is not significantly tuned in experiment or in simulation. Thus, we can conclude that the multiphysics simulation approach accounting for the thermodynamic energy minimization of the LC alignment under a spatially varying external electrostatic field developed in this work provides a more accurate description of the expected transmittance when compared to the assumption of uniform aligned medium of LCs.

**Alignment of liquid crystals around a periodic metasurface nanostructure array**

Next, we take a closer look at how LCs align in a periodic array of nanostructures to gain a better physical understanding of our device. **Figure 3** plots simulation results for each step of our multiphysics modeling used in **Fig. 2**. When there is no applied voltage (**Figs. 3a-d**), we first simulate the LC director field using the Landau-de Gennes equations. The resulting director field profile is then converted to a spatially varying diagonal optical refractive index which is imported into Lumerical to simulate the final optical transmittance spectrum and near field optical electric field (**Figs. 3b,d**). To model the dynamic response of our device (**Figs. 3e-j**), we first calculate the spatially varying diagonal DC dielectric permittivity from the LC director field simulated at 0 V (**Figs. 3a,c**) and use this to simulate the spatially varying electrostatic field throughout the cell using finite-element-method simulations in COMSOL (**Figs. 3e,h**). We used ordinary and extraordinary DC permittivities of E7 of 6 and 17, respectively [15]. Additionally, we used a DC permittivity of 100 for $TiO_2$ [36]. Then, we recalculated the LC director field (**Figs. 3f,i**) using Landau-de Gennes simulations under the spatially varying external field shown in **Figs. 3e,h**. Finally, we use the resulting refractive index profile to calculate the final transmittance and near field optical electric field profiles (**Figs. 3g,j**). Effectively, this segregated, sequential simulation approach provides some level of coupling between LCs and the optical field from voltage application. Fully coupled approaches will be considered in future work.

By analyzing the results for the 0 V case in **Fig. 3**, we can see that LCs at the top of the cell (nearest the alignment layer at z = 2000 nm) are aligned along the y-axis, as expected. As we move down the cell, towards the $TiO_2$ pillar, there are some non-uniformities that may represent a local energy minimum of LCs, however the LCs remain mostly aligned along the y-axis. Near the $TiO_2$ nanopillar, the LCs begin to be more strongly disturbed and point along the normal of the pillar walls. This alignment along the normal is due to the homeotropic anchoring condition of the pillars implemented in the simulation and enforced experimentally by the HMDS surface coating. The optical mode at $\lambda$ = 965 nm (**Figs. 3b,d**) shows that the optical field has a strong enhancement in the $TiO_2$ pillar ~ 500 nm above and laterally between neighboring pillars.

Next, we investigate how the LCs and optical fields change under an applied electrostatic field. We first plot the spatially varying electrostatic field profile (**Figs. 3e,h**). The field is concentrated just above the $TiO_2$ pillar and very little field is in between adjacent pillars. This is a result of $TiO_2$ having a much higher DC permittivity than that of the LCs (ordinary and extraordinary) which causes most of the voltage to drop across the LC from z = 310 – 2000 nm. Because of the large permittivity mismatch between the LCs and $TiO_2$, this also means that the arrangement of the LCs has a limited effect on the distribution of electric field throughout the cell. To more uniformly distribute the electric field throughout the entire cell (especially in between the pillars), one could choose different materials with more closely matched permittivities for the nanostructures and LCs. For example, $SiO_2$ has a much lower DC permittivity and other LCs such as dual-frequency LCs have higher DC permittivities, reaching values above 100 [37]. The resulting director fields plotted in **Figs. 3f,i** show the importance of considering this spatially varying electric field. At the top of the cell, we again see that the LCs prefer to align along the y-axis given by the alignment layer and gradually rotate along the z-axis lower into the cell. In between the $TiO_2$ pillars, however, there is minimal change in the LC director field compared to the 0 V case. This is because the electrostatic field is minimal in this region of the cell. As a result, the LCs directly above the $TiO_2$ pillar experience the largest rotation due to the applied field while those in between the pillars experience near-zero rotation. While the amount of LC tuning strongly depends on materials selection and device architecture, this finding is important as many previous studies have assumed that the LCs switch uniformly around the metasurface nanostructures. Such an assumption could result in discrepancies between simulation and experiment. This result reinforces the importance of considering the spatially varying alignment of LCs at the nanoscale.

Focusing on the simulated optical fields, we can see the final effect of the LC reorientation. For the mode of interest at $\lambda$ = 965 nm, **Figs. 3b,d,g,j** show that the optical field profiles in both YZ and XY planes are significantly changed between the 0 V and 10 V simulations. This change in optical field is primarily driven by the rotation of LCs directly above the pillar as this is where there was the greatest overlap of optical field and LC rotation. If the LCs were also switched in between the pillars, we would expect even more optical modulation. Given this finding, future optical metasurface designs using the same materials presented here could be considered which are most sensitive to refractive index modulation above the nanostructure to optimize metasurface response. This redesign could potentially be implemented easier than changing LC and nanopillar materials to force LCs to switch more uniformly.

A more detailed look at the measured voltage-dependent modulation of select active metasurfaces with diameters from, D = 280-480 nm, is provided in Supplementary Information (Fig. S2). This provides the measured transmittance for intermediate voltage values from those presented in **Fig. 2**. Here, we can more clearly see that the largest modulation is achieved for structures with intermediate pillar widths. Pillars with diameters of approximately 360 - 440 nm see the largest modulation for the optical mode centered around 940 – 960 nm. Thus, for our final demonstration, we fabricated a large area metasurfaces (5x5 $mm^2$) with circular cylindrical pillars of diameter 400 nm and used a top substrate consisting of the photoconductor, BSO, to demonstrate 2D optically addressed patterning of the polarization rotation of a laser beam.

**Large-area, highly efficient patterning of polarization in two dimensions**

Here, we present the results of our large area (5x5 mm$^2$) metasurface used in the experimental configuration shown in **Fig. 1d**. The optical addressing of multiple patterns with features on the mm-scale down to 10s of $\mu$m is shown in **Fig. 4a**. The Caltech logo in (i) is 1.5 mm across, while the vertical bars in (v) have a width and spacing of 20 $\mu$m, approximating the resolution limit of our device. The polarization azimuth and ellipticity of the transmitted light is plotted in **Fig. 4b** when the entire area is illuminated with blue light. Here, we see that the azimuth is rotated by ~ 90° between an applied voltage of 0 and 20 V$_{p-p}$ while maintaining a relatively small ellipticity. In **Fig. 4c**, the total transmittance is plotted as a function of voltage showing a minimum and maximum transmittance of ~ 18% and > 60%, respectively. Thus, in **Fig. 4a**, while modulating with 20 V$_{p-p}$, the bright areas have a transmittance of ~ 60% (20 V$_{p-p}$) while the dark areas initially had a transmittance of ~ 30% (0 V$_{p-p}$). However, in our images, the dark areas have a transmittance of near zero because of our cross-polarizer arrangement and low ellipticity (~ -0.2) at 0 V. This means that we can transmit polarization rotated light with an arbitrary 2D pattern with an efficiency > 60%.

To test the speed benefits of the thinner LC layer in our device, we compared the time-response of our thinner LC-metasurface device to a device with only LCs and a thicker liquid crystal gap. Since the device with only LCs has no optical resonances, the LC layer must be much thicker in order to accumulate sufficient phase and achieve the same 90° polarization conversion. The LC layer thickness of this metasurface-free device was ~ 6 $\mu$m. **Figures 4d,e** show the normalized transient transmittance when the voltage is turned on and off, respectively. The characteristic ON and OFF times, $\tau_{ON}$ and $\tau_{OFF}$, to reach 90% and 10% of the maximum transmittance values, respectively, are shown as insets. The purely LC cell turns on ($\tau_{ON}$ = 10.8 ms) faster than the LC-metasurface cell ($\tau_{OFF}$ = 20.8 ms) which is likely caused by stronger anchoring in the LC-metasurface, which requires larger voltages to drive the LCs in the 'ON' state ($\tau_{ON} \propto 1/V^2$). However, when the voltage is turned 'OFF' and there is no longer a driving voltage, the LC-metasurface cell switches faster ($\tau_{OFF}$ = 16.4 ms) than the non-resonant LC cell ($\tau_{OFF}$ = 44.0 ms) because of the thinner LC layer. It is of interest that the LC-metasurface cell turns off faster than it turns on. This is an atypical behavior, likely caused by the complex LC-dynamics and anchoring to the TiO$_2$ nanostructures. By comparing the slowest response time for each LC cell ($\tau_{OFF}$ = 44.0 ms for the purely LC cell and $\tau_{ON}$ = 20.8 ms for the LC-metasurface cell), we conclude that the LC-metasurface improves the switching speed by a factor of approximately two.

## 3. Discussion

We have demonstrated a metasurface-enhanced optically-addressed SLM for the 2D patterning of transmitted polarization conversion of near-IR light with > 60% efficiency which is suitable for high-power laser illumination. Our device demonstrates a two-fold improvement in switching speed over a similar device architecture which uses only LCs and no metasurface enhancement. In addition, we introduce a multiphysics approach to modeling the optical response of LC-based active nanophotonic devices. This method involves solving the complex spatially varying LC orientation under an external electrostatic field and simulating the final transmitted optical field through this cell. The experimental and simulation results presented in this work pave the way for the design of more complex LC-based active nanophotonic devices.

Despite the maturity of LC-based active optical devices, the orientation of LCs on the nanoscale around nanostructured features has remained an outstanding problem. This work takes a step towards solving these complex interactions by providing a simulation framework as well as detailed experimental measurements. Despite the gains presented here, there is still a gap between our model and experiment. This mismatch is most likely a result of many unknown material values which must be assumed in our simulation models. For example, the anchoring strengths of the LCs on each surface of our device (i.e. top polyimide alignment layer, $TiO_2$ pillars, bottom $SiO_2$ surface) were assumed based on literature values. (See Supplementary Information for more details of all simulation inputs and assumptions.) While the anchoring strength of LCs on each surface can be measured for planar surfaces, it is much harder to predict these values on surfaces such as our $TiO_2$ nanopillars which have nanoscale features and roughness. This complex geometry may be altering the surface chemistry in unexpected ways. Performing more detailed measurements of the fundamental properties of LCs (such as anchoring strength) on surfaces of different materials with varying surface roughness and surface coatings may elucidate some trends which could be incorporated into future works to improve the LC modeling. While these measurements may provide some insight, it is always difficult to exactly reproduce the conditions which occur in a final active SLM device. Thus, this highlights the importance of collecting detailed optical measurements such as those presented here. As our resonant metasurfaces are most sensitive to the LC orientation in a length scale of ~ 100s nm around each $TiO_2$, we are effectively probing these complex LC interactions by measuring the transmittance or polarization spectrum as a function of applied voltage. While this work has focused on the modulation of near-IR light in free-space, the methods and results presented here can be applied more generally to any active optical device using LCs. For example, many integrated photonic devices such as optical phased arrays use LCs as active phase shifting elements. Even though waveguide geometries are often larger and simpler than the $TiO_2$ nanopillars presented in this work, the nanoscale alignment of LCs in the 100s of nm surrounding the waveguide must still be precisely modeled. Thus, the conclusions from this work can be applied to a wide range of optical devices using active LC tunability.

## 4. Materials and methods

**Sample fabrication**

We first start with 40x40 mm$^2$ sapphire substrates coated with thermally treated indium-tin-oxide (ITO) on the top surface, and with an anti-reflection coating on the bottom surface. These are the substrates used by Seurat Technologies for their existing non-resonant optically addressed light valves (OALVs) for metal 3D printing. These substrates are shipped to Caltech where we first prepare the substrates with a 5 minute $O_2$ plasma clean. Then, ~ 700 nm $SiO_2$ is deposited via plasma-enhanced chemical vapor deposition (PECVD) using an Oxford Instruments Plasma Technology Plasmalab System 100 system, processed at 200 °C. The deposited films have an ellipsometrically measured and fitted n ~ 1.45 and k ~ 0 at $\lambda$ = 1 $\mu$m when using a Sellmeier model. Next, we deposit our ~ 300 nm $TiO_2$ film using pulsed DC reactive sputtering at room temperature in an AJA UHV Orion dielectric sputter system. By sputtering a pure Ti target in a Ar/$O_2$ environment, we obtain high quality $TiO_2$ on our substrate. The $O_2$ flow rate, chamber pressure, and sputtering power was optimized to obtain a $TiO_2$ film with high refractive index, low loss, and relatively high deposition rate. Our deposition parameters are: 100 kHz, 3 $\mu$s, and 200 W for the DC power source, a flowrate of 20 sccm Ar and 2 sccm $O_2$ to the main chamber while maintaining a process pressure of 4.5 mTorr. We find these conditions provide a deposition rate of ~ 5 nm/min. After deposition, we anneal our films at 180 °C in air for 10 minutes which likely increases the oxygen content of our films, resulting in lower loss. We obtained an ellipsometrically measured and fitted n ~ 2.3 and k ~ 0 at $\lambda$ = 1 $\mu$m using a Tao-Lorentz model. Next, we deposited 50 nm Cr at a rate of 0.5 Å/s using an AJA International ATC Orion Series system electron beam (e-beam) evaporator. This layer will be patterned and used as a hard mask for our $TiO_2$ etch. Next, we prepared our sample for e-beam lithography (EBL) by spinning with acetone and isopropyl alcohol (IPA), baking at 180 °C for 10 minutes, then another 5 minute $O_2$ plasma clean. This step improves the adhesion of our e-beam resist, ma-N 2403. After the $O_2$ clean, we used a spin coater to spin ma-N 2403 at 3000 rpm for 1 minute and baked at 90 °C for 1 minute. This should give a film of ~ 300 nm. We then wrote our metasurface features using a Raith electron beam pattern generator (EBPG) 5200: 100 kV Electron Beam Lithography tool with a beam current of 5 nA and dose of 450 $\mu C/cm^2$. After exposure, we developed our sample in MF-319 for 1 minute and rinsed in water. Next, we used an Oxford Instruments Plasma Technology Plasmalab System 100 ICP-RIE 380 system to first etch into the Cr layer using our ma-N as a mask and $Cl_2$ chemistry. We then switched to a $SF_6/C_4F_8$ chemistry to etch into our $TiO_2$ using the Cr as a hard mask. For the Cr etch, we etched for 2 minutes using the following parameters: chamber pressure = 15 mTorr, ICP power = 1200 W, RF power = 60 W, substrate temperature = 15 °C, $Cl_2$ flowrate = 54 sccm, $O_2$ flowrate = 4 sccm. Then, we etched into the $TiO_2$ for 3 minutes using the following parameters: 2 mTorr chamber pressure, ICP power = 1000 W, RF power = 150 W, substrate temperature = 20 °C, $C_4F_8$ flowrate = 15 sccm, $SF_6$ flowrate = 6 sccm, Ar flowrate = 5 sccm. All these parameters were optimized to obtain a good quality etch with near-vertical sidewalls. (It should be noted that we performed multiple $SF_6$ and $O_2$ cleans before and after each step in order to clean and prepare the etcher chamber.) Finally, after etching, we stripped any remaining Cr by submerging our sample in CR-7S Cr etchant for 5 minutes. At this point, we have a passive $TiO_2$ metasurface which was extensively characterized, then shipped to Seurat Technologies for integration of LCs.

Substrates were prepared for filling with LCs by spin coating a layer of polyimide on the top (non-metasurface) substrate and mechanically rubbing along the y-axis to create an alignment layer. The two substrates were then bonded together with a fixed spacing set by microspheres of diameter ~ 2 $\mu$m dispersed in an epoxy. After the gap was set and the epoxy was cured, the LCs were flowed into the gap and then the device was sealed. Finally, electrical contacts were placed on each substrate and the final devices were shipped back to Caltech for optical characterization.

**Measurements**

All measurements were performed using a home-built optical setup depicted in **Fig. 1d**. To measure transmittance through our sample, a beamsplitter was placed after the second lens on the IR path and the light was collected on a photodiode. This light was used as a reference to remove unwanted noise from intensity fluctuations in the laser. Another beamsplitter and photodiode was placed just in front of the camera to collect the light transmitted through the sample. In one measurement, the power after the metasurface was divided by the power before the metasurface to obtain the relative transmitted power. This was repeated with the metasurface removed from the setup, to approximate a sample transmittance of 1. Our sample measurement was then divided by this "air" measurement to obtain the final transmittance values presented. Note: All data presented in **Fig. 2** was performed with the final cross-polarizer removed and no blue light illumination.

**Simulations**

All LC-based Landau-de Gennes simulations were performed using the open-source python code, open-Qmin, developed as a collaboration between the research groups of Daniel Sussman (Emory University) and Daniel Beller (Johns Hopkins University) [23]. The Supplementary Information provides more details about the open-Qmin simulation setup and assumptions. All electrostatic simulations were performed using finite-element-method (FEM) in COMSOL Multiphysics electrostatic module. All optical domain simulations were performed using Ansys Lumerical finite-difference time-domain (FDTD) Solutions.

## 5. Back matter

5.1 Funding

This work was supported by the Air Force Office of Scientific Research Meta-Imaging MURI Grant No. FA9550-21-1-0312 (J.S., C.U.H., R.S., and H.A.A.). J.S. acknowledges the support of the Natural Sciences and Engineering Research Council of Canada (NSERC) through the Postgraduate Scholarship – Doctoral program. J.S., C.U.H., R.S., and H.A.A. gratefully acknowledge support from Seurat Technologies.

5.2 Acknowledgements

We gratefully acknowledge the critical support and infrastructure provided for this work by The Kavli Nanoscience Institute at Caltech. The authors would also like to thank Phillippe Pearson for fabricating the chrome shadow-mask used for projecting images on our active device.

5.3 Disclosures

The authors declare no conflicts of interest

5.4 Author Contributions

J.S., C.U.H., Z.S.D., R.S., S.E., and H.A.A. conceived the ideas for this research project. J.S., with assistance and input from C.U.H., performed the optical design, fabricated the $TiO_2$ metasurfaces, and performed all low-power optical measurements. Z.S.D., A.M.K.F., and J.G. integrated liquid crystals to form final devices and provided input on design considerations. A.M.K.F. performed high-power damage-threshold measurements. R.S. provided input on design considerations throughout the project. All authors contributed to the writing of the manuscript.

**References**

1. Li, L. *et al.* Electromagnetic reprogrammable coding-metasurface holograms. *Nat Commun* **8**, 197 (2017).
2. Zhang, L. *et al.* Space-time-coding digital metasurfaces. *Nat Commun* **9**, (2018).
3. Li, Q. *et al.* High-efficiency broadband active metasurfaces via reversible metal electrodeposition. *Light Sci Appl* **15**, 38 (2026).
4. Shirmanesh, G. K., Sokhoyan, R., Wu, P. C. & Atwater, H. A. Electro-optically Tunable Multifunctional Metasurfaces. *ACS Nano* **14**, 6912–6920 (2020).
5. Wu, C. *et al.* Programmable phase-change metasurfaces on waveguides for multimode photonic convolutional neural network. *Nat Commun* **12**, 96 (2021).
6. Fukui, T. *et al.* 17-GHz lossless InP-membrane active metasurface. *Sci Adv* **11**, (2025).

7. Dudley, D., Duncan, W. M. & Slaughter, J. Emerging digital micromirror device (DMD) applications. in (ed. Urey, H.) 14 (2003). doi:10.1117/12.480761.
8. Zhang, Z., You, Z. & Chu, D. Fundamentals of phase-only liquid crystal on silicon (LCOS) devices. *Light Sci Appl* **3**, e213–e213 (2014).
9. Engström, D., O'Callaghan, M. J., Walker, C. & Handschy, M. A. Fast beam steering with a ferroelectric-liquid-crystal optical phased array. *Appl Opt* **48**, 1721 (2009).
10. Li, S.-Q. *et al.* Phase-only transmissive spatial light modulator based on tunable dielectric metasurface. *Science (1979)* **364**, 1087–1090 (2019).
11. Moitra, P. *et al.* Electrically Tunable Reflective Metasurfaces with Continuous and Full-Phase Modulation for High-Efficiency Wavefront Control at Visible Frequencies. *ACS Nano* **17**, 16952–16959 (2023).
12. Carbajo Garcia, S. & Bauchert, K. Power handling for LCoS spatial light modulators. in *Laser Resonators, Microresonators, and Beam Control XX* (eds. Kudryashov, A. V., Paxton, A. H. & Ilchenko, V. S.) 64 (SPIE, 2018). doi:10.1117/12.2288516.
13. Mansha, S. *et al.* High resolution multispectral spatial light modulators based on tunable Fabry-Perot nanocavities. *Light Sci Appl* **11**, 141 (2022).
14. Haiying Wang. Studies of Liquid Crystal Response Time. (University of Central Florida, 2005).
15. Maruthiyodan Veetil, R. *et al.* Nanoantenna induced liquid crystal alignment for high performance tunable metasurface. *Nanophotonics* **0**, (2023).
16. Sun, M. *et al.* Efficient visible light modulation based on electrically tunable all dielectric metasurfaces embedded in thin-layer nematic liquid crystals. *Sci Rep* **9**, 8673 (2019).
17. Zhu, Z. *et al.* Metasurface-enabled polarization-independent LCoS spatial light modulator for 4K resolution and beyond. *Light Sci Appl* **12**, 151 (2023).
18. Izdebskaya, Y. V., Yang, Z., Shvedov, V. G., Neshev, D. N. & Shadrivov, I. V. Multifunctional Metasurface Tuning by Liquid Crystals in Three Dimensions. *Nano Lett* **23**, 9825–9831 (2023).
19. Dolan, J. A. *et al.* Broadband Liquid Crystal Tunable Metasurfaces in the Visible: Liquid Crystal Inhomogeneities Across the Metasurface Parameter Space. *ACS Photonics* **8**, 567–575 (2021).
20. van Heijst, E. A. P. *et al.* Electric tuning and switching of the resonant response of nanoparticle arrays with liquid crystals. *J Appl Phys* **131**, (2022).
21. Atorf, B., Mühlenbernd, H., Muldarisnur, M., Zentgraf, T. & Kitzerow, H. Effect of Alignment on a Liquid Crystal/Split-Ring Resonator Metasurface. *ChemPhysChem* **15**, 1470–1476 (2014).
22. De Cort, W., Beeckman, J., Claes, T., Neyts, K. & Baets, R. Wide tuning of silicon-on-insulator ring resonators with a liquid crystal cladding. *Opt Lett* **36**, 3876 (2011).
23. Xue, X., Nys, I., Neyts, K. & Beeckman, J. Influence of period and surface anchoring strength in liquid crystal optical axis gratings. *Soft Matter* **18**, 3249–3256 (2022).
24. Sussman, D. M. & Beller, D. A. Fast, Scalable, and Interactive Software for Landau-de Gennes Numerical Modeling of Nematic Topological Defects. *Front Phys* **7**, (2019).
25. Tumkur, T. U. *et al.* Toward high laser power beam manipulation with nanophotonic materials: evaluating thin film damage performance. *Opt Express* **29**, 7261 (2021).
26. Sokhoyan, R. *et al.* Electrically tunable conducting oxide metasurfaces for high power applications. *Nanophotonics* **12**, 239–253 (2023).

27. Taghizadeh, M. & Zhu, Z. H. A comprehensive review on metal laser additive manufacturing in space: Modeling and perspectives. *Acta Astronaut* **222**, 403–421 (2024).
28. Manzur, T. Free Space Optical Communications (FSO). in *2007 IEEE Avionics, Fiber-Optics and Photonics Technology Conference* 21–21 (IEEE, 2007). doi:10.1109/AVFOP.2007.4365728.
29. Brown, W. C. The technology and application of free-space power transmission by microwave beam. *Proceedings of the IEEE* **62**, 11–25 (1974).
30. Antman, Y. *et al.* High power chip-scale laser. *Opt Express* **32**, 47306 (2024).
31. Elhadj, S., Davidson, Z. & Sargol, Y. A light-driven light valve for metal additive manufacturing. in *Smart Materials for Opto-Electronic Applications* (eds. Rendina, I., Petti, L., Sagnelli, D. & Nenna, G.) 3 (SPIE, 2023). doi:10.1117/12.2665533.
32. Jacobsen, A. D., Beard, T. D., Bleha, W. P., Margerum, J. D. & Wong, S. Y. The liquid crystal light valve, an optical-to-optical interface device. in *NASA. Goddard Space Flight Center Proc. of the Conf. of Parallel Image Process. for Earth Observation Systems* (1972).
33. Seurat Technologies. How Area Printing Works. *https://www.seurat.com/area-printing* (2025).
34. Chatterjee, B. *et al.* Optically Addressable Light Valve Based on a GaN:Mn Photoconductor. *ACS Applied Engineering Materials* **3**, 325–336 (2025).
35. Buckley, B. W. *et al.* Beam-based programmable spatial shaper for reducing optics exchange in the National Ignition Facility. in *High Power Lasers for Fusion Research VIII* (eds. Häfner, C. L. & Awwal, A. A.) 13 (SPIE, 2025). doi:10.1117/12.3043952.
36. Manyuhina, O. V., Tordini, G., Bras, W., Maan, J. C. & Christianen, P. C. M. Doubly periodic instability pattern in a smectic-A liquid crystal. *Phys Rev E* **87**, 050501 (2013).
37. Bonkerud, J., Zimmermann, C., Weiser, P. M., Vines, L. & Monakhov, E. V. On the permittivity of titanium dioxide. *Sci Rep* **11**, 12443 (2021).
38. Hsiao, C.-C. & Lee, W. Dielectric characterization of dual-frequency liquid crystals doped with cetyltrimethylammonium bromide. *J Mol Liq* **439**, 128933 (2025).

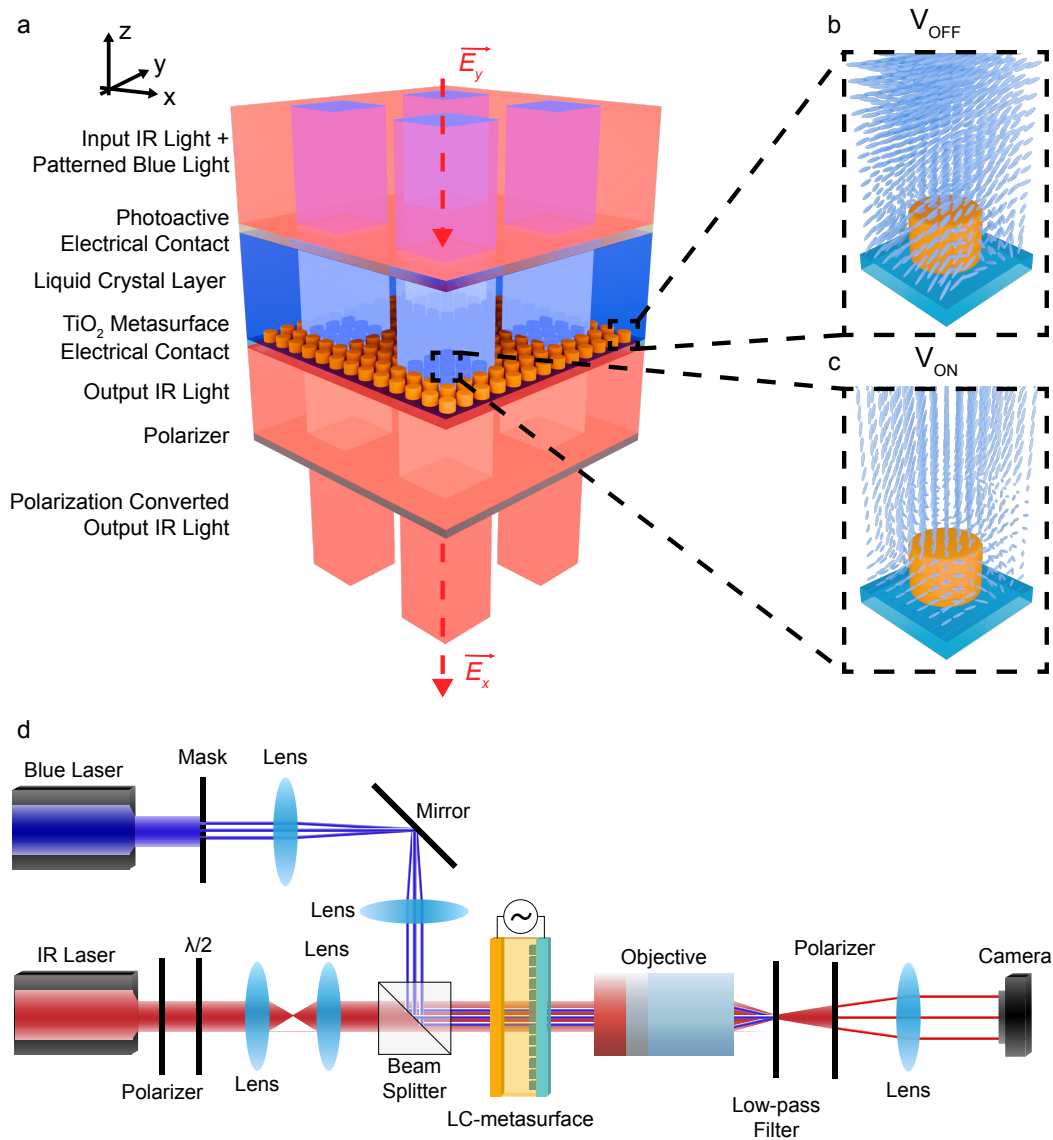

**Figure 1: An LC-based metasurface-enhanced optical light valve. (a)** Schematic of a liquid crystal (LC)-actuated active metasurface producing the linear polarization conversion of $\lambda \sim 1~\mu m$ polarized IR "probe" light using a patterned (masked) blue ($\lambda = 435$ nm) photoexcitation laser as the addressable patterning "write" light. **(b),(c)** Magnified views of (a) showing the LC arrangement around individual TiO$_2$ resonators in regions where the blue laser does not illuminate (no field transfer to LC) and does illuminate (field transferred to LC), respectively. **(d)** Schematic of experimental optical setup of our metasurface enhanced optical light valve.

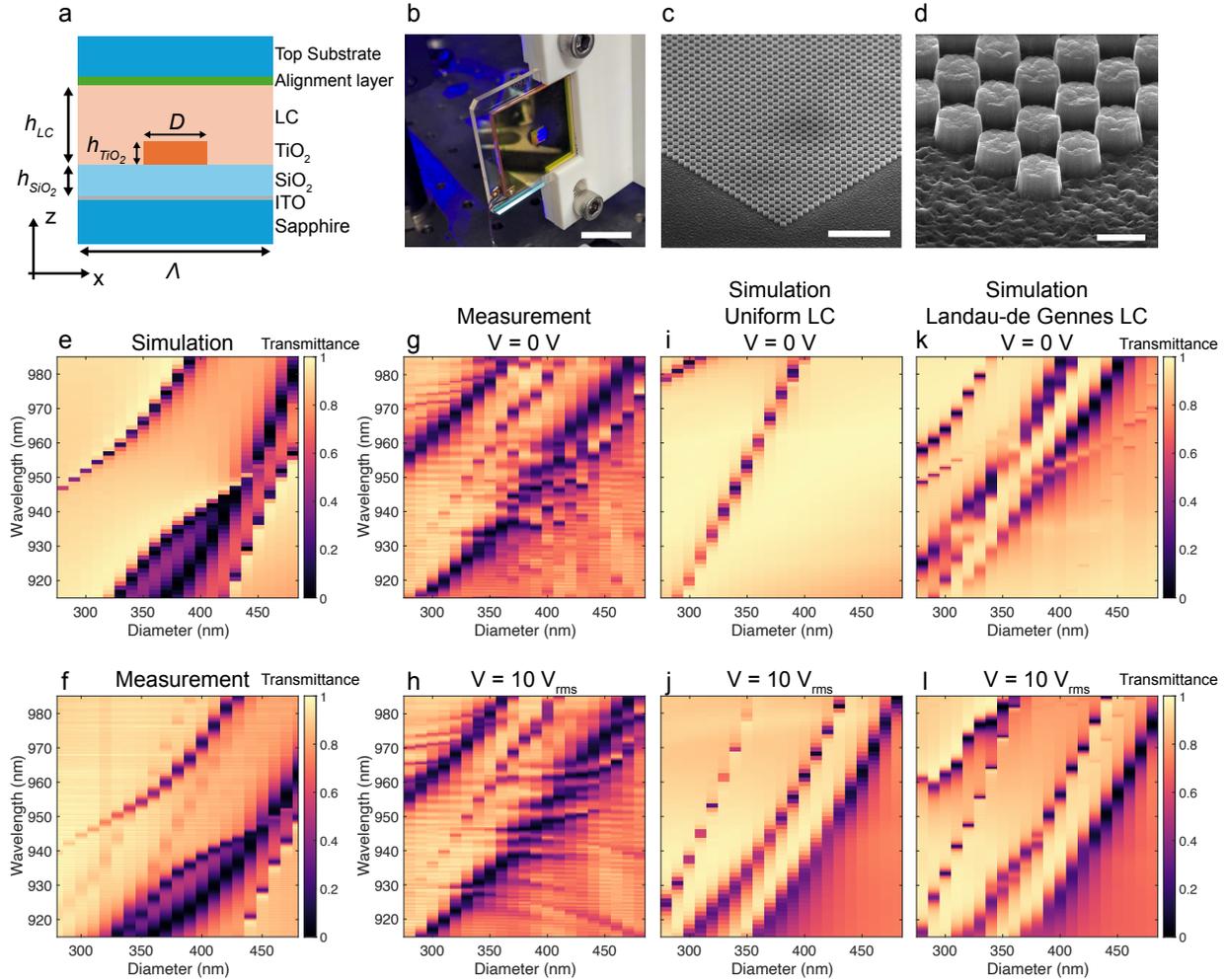

**Figure 2**: **A liquid crystal (LC)-actuated TiO$_2$ metasurface. (a)** Cross-section schematic of device unit cell where $h_{TiO_2}$ = 310 nm, $h_{LC}$ = 2 μm, $h_{SiO_2}$ = 700 nm, $D$ = 280−480 nm, and $\Lambda$ = 574 nm (spacing between elements). The alignment layer is rubbed in the y-axis (into the plane). For all transmittance measurements in this figure, the top substrate is ITO on sapphire. **(b)** Optical image of the full device in our optical setup, illuminated with the IR and blue photoexcitation lasers. Scale bar: ~2 cm. **(c),(d)** Tilted scanning electron micrographs of the TiO$_2$ metasurface on SiO$_2$. Scale bars: 5 μm and 500 nm, respectively. **(e),(f)** Simulated and measured, respectively, transmittance spectra of metasurfaces with varying TiO$_2$ pillar diameters embedded in 1.5 μm poly-methyl methacrylate (PMMA). **(g),(h)** Transmittance measurement of the metasurface, filled with LCs, with 0 V and 10 V$_{rms}$ applied voltage, respectively. **(i),(j)** Simulated transmittance of metasurface, filled with LCs, with 0 V and 10 V$_{rms}$ applied voltage, respectively, assuming uniform distribution of LC orientation. **(k),(l)** Simulated transmittance of metasurface, filled with LCs, with 0 V and 10 V$_{rms}$ applied voltage, respectively, using LC orientation calculated from Landau-de Gennes theory [23].

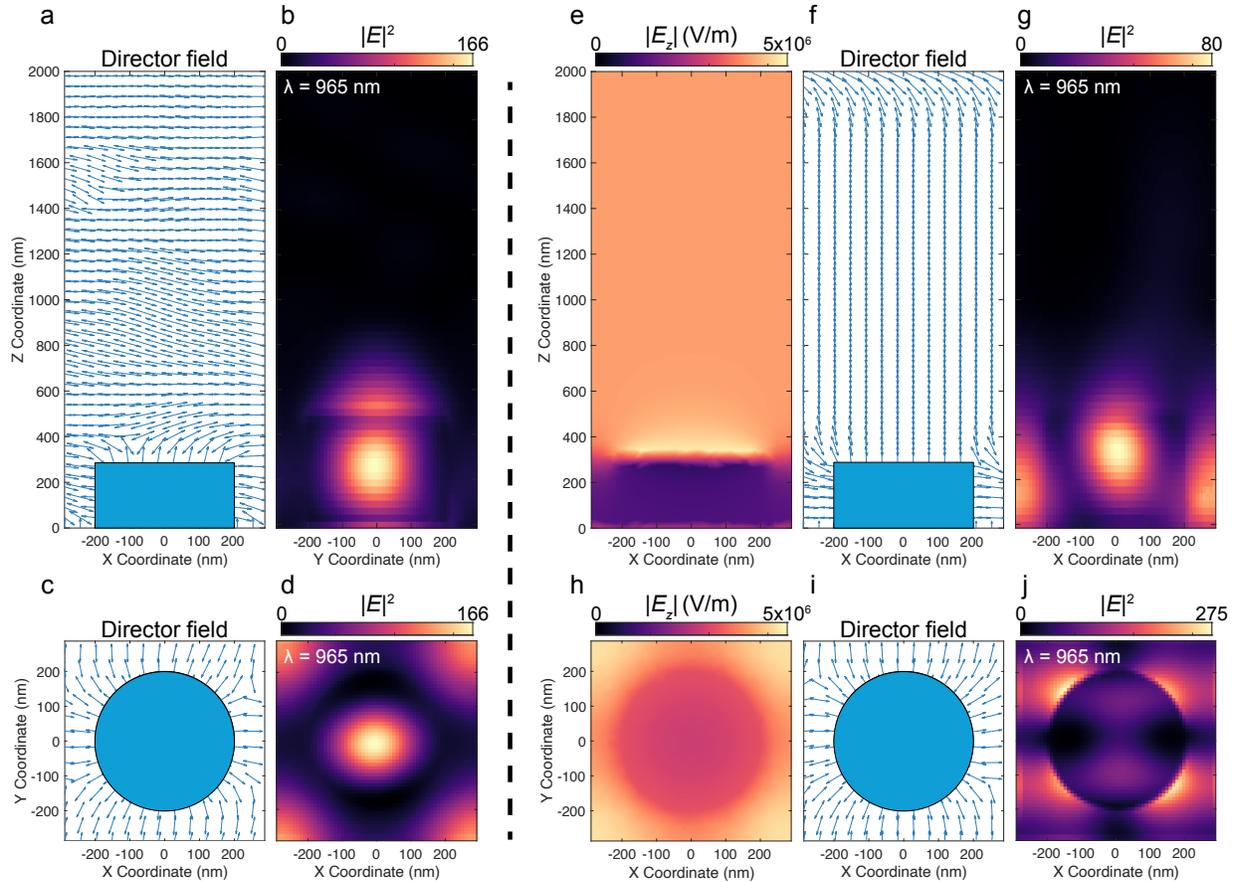

**Figure 3: Voltage-dependent liquid crystal (LC) arrangement around TiO₂ nanostructures.** **(a)–(d)** Simulated LC director fields (a and c) and optical field profiles (b and d) for YZ and XY cross-sections of a single metasurface unit cell with no applied bias (0 V). **(e)-(j)** Simulated electrostatic field profiles (e and h), LC director fields (f and i), and optical field profiles (g and j) for YZ and XY cross-sections of a metasurface unit cell with an applied bias of 10 V. The blue overlayed rectangles and circles represent the simulated TiO₂ pillar which has a diameter of 400 nm.

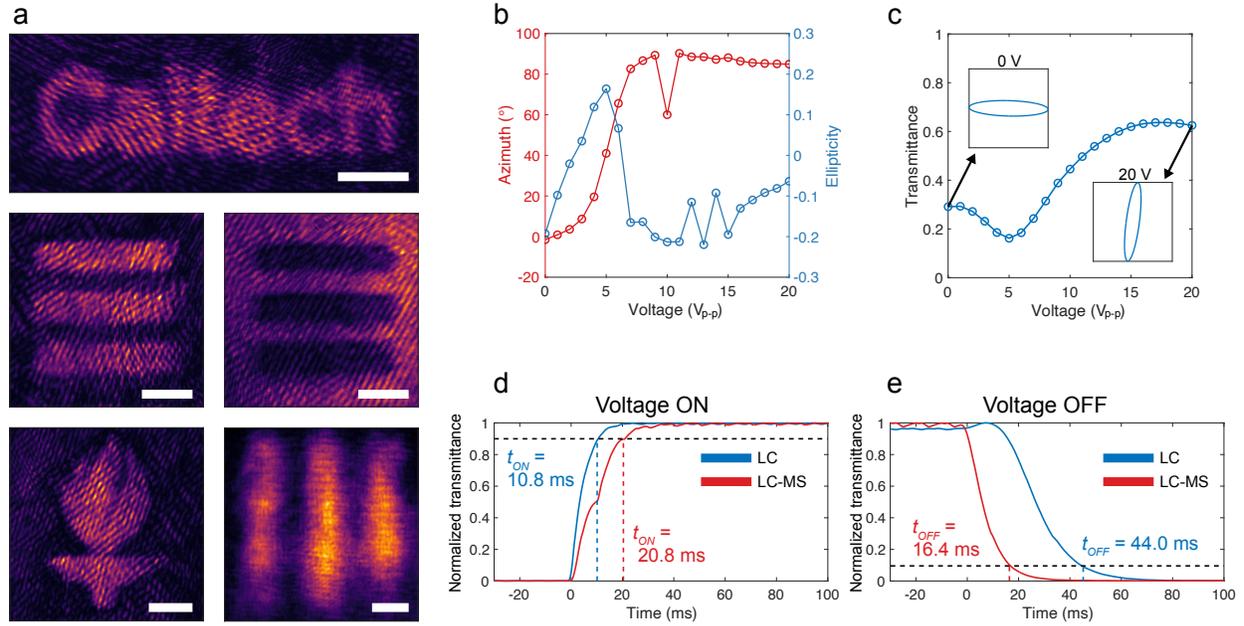

**Figure 4: 2D control of polarization converted light spanning μm- to mm-scale. (a)** Images of spatial patterning of IR beam polarization. All images, except (iii), use a cross-polarizer arrangement. (iii) Uses a parallel polarizer arrangement. Scale bars for (i)–(v) are 300 $\mu m$, 250 $\mu m$, 250 $\mu m$, 250 $\mu m$, and 20 $\mu m$, respectively. **(b)** Measured azimuth and ellipticity of transmitted light as a function of voltage. **(c)** Measured transmittance as function of voltage. Insets at 0 V and 20 $V_{p-p}$ plot the transmitted polarization ellipse for each voltage. **(d),(e)** Measured time-response (normalized photodiode response on y-axis) for the metasurface-free (blue line) and metasurface device (red line) when the voltage is turned on and off, respectively. The applied voltage was switched between 0 and 20 $V_{p-p}$ and the incident wavelength was $\lambda = 962.2$ nm for all time-traces. The measured characteristic ON and OFF times, $\tau_{ON}$ and $\tau_{OFF}$, are shown as insets.

Page 17 of 23

Supplementary Information for

# An Optically Addressable Transmissive Liquid Crystal Metasurface Spatial Light Modulator

Jared Sisler[1], Claudio U. Hail[1], Zoey S. Davidson[2], Austin M. K. Fehr[2], Jiannan Gao[2], Ruzan Sokhoyan[1], Selim Elhadj[2], and Harry A. Atwater[1]

*[1] Thomas J. Watson Laboratories of Applied Physics, California Institute of Technology, Pasadena, California 91125, USA*

*[2] Seurat Technologies Inc., Wilmington, Massachusetts 01887, USA*

Corresponding authors:
Selim Elhadj and Harry A. Atwater

**Content**

Analysis of optical modes in passive $TiO_2$ metasurface
Measurements of transmittance modulation of active metasurface devices with varying dimensions
Description of liquid crystal simulations

**Fig. S1:** Optical field profiles of simulated TiO2 metasurface submerged in PMMA
**Fig. S2:** Measured transmittance modulation of LC-infiltrated TiO2 metasurfaces as a function of voltage.

## Optical mode profiles for a passive TiO$_2$ metasurface in a PMMA environment

Here, we present the simulated optical electric and magnetic field profiles (Fig. S1) corresponding to the simulated transmittance modes plotted in Fig. 2e of the main text. In the simulation, we assume a structure of a 310 nm tall TiO$_2$ cylindrical pillar with a diameter of 400 nm which is periodic in the x- and y-axes with a period of 574 nm in each axis. The TiO$_2$ metasurface (n = 2.3) sits on a 700 nm continuous film of SiO$_2$ (n = 1.46) which is on top of a sapphire substrate (n = 1.75). The TiO$_2$ metasurface is submerged in a planarized 1.5 $\mu$m thick film of poly-methylmethacrylate (PMMA) (n = 1.49). The structure is illuminated from above with linearly polarized light along the Y-axis and a 1° off-axis angle of incidence in the x- and y-axes.

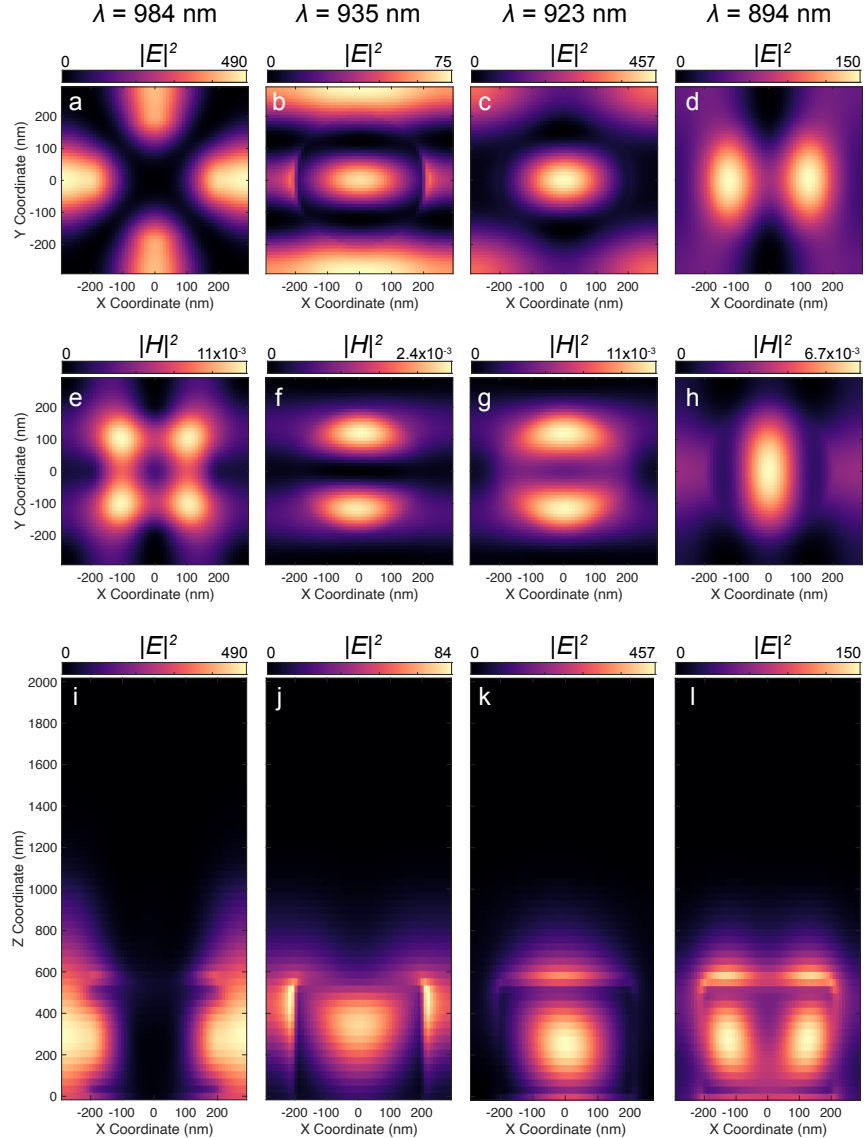

**Figure S1. Optical field profiles of simulated TiO$_2$ metasurface submerged in PMMA.** All plots correspond to a TiO$_2$ pillar of height = 310 nm and diameter = 400 nm at four wavelengths: $\lambda$ = 984 nm **(a),(e),(i),** $\lambda$ = 935 nm **(b),(f),(j),** $\lambda$ = 923 nm **(c),(g),(k),** and $\lambda$ = 894 nm **(d),(h),(l).** (a)-(d) Plotted electric field profiles in the XY plane (Z = 155 nm). (e)-(h) Plotted magnetic field

profiles in the XY plane (Z = 155 nm). (i)-(l) Plotted electric field profiles in the XZ plane (Y = 0).

By analyzing the mode profiles in Fig. S1, we can conclude that the modes at $\lambda$ = 935 nm and $\lambda$ = 923 nm are predominantly electric dipole modes concentrated within the TiO$_2$ pillar, the mode at $\lambda$ = 894 nm is a magnetic dipole concentrated in the TiO$_2$ pillar, and the mode at $\lambda$ = 985 nm is a higher order mode concentrated outside of the TiO$_2$ pillar. The mode at $\lambda$ = 935 nm has significant electric field distribution outside of the pillar which is delocalized in the x-axis. This is likely the cause for this mode to exhibit different dispersion from the mode at $\lambda$ = 923 nm which is similar in character but more concentrated to the TiO$_2$ pillar.

Due to our excellent agreement between simulation and experiment, we conclude that these are the same modes which we measure in our device.

**Voltage-dependent transmittance of LC-integrated active metasurfaces**

Here, we present (Fig. S2) the measured voltage-dependent transmittance spectra for multiple LC-metasurfaces with a uniform ITO-sapphire top substrate. The devices presented here are those presented in Fig. 2 of the main text. From the measured transmittance modulation, we can see multiple optical modes which are tuned depending on the metasurface dimensions and mode. In general, metasurfaces with smaller $TiO_2$ diameters (280 – 320 nm) exhibit minimal modulation as the driving voltage amplitude is increased. The maximum modulation is seen for intermediate pillar diameters (~ 400 nm) and there are many different modes which are all tuned simultaneously. Thus, our large-area polarization patterning demonstrated in Fig. 5 of the main text used a 5x5 mm² metasurface consisting of $TiO_2$ pillars with a diameter of 400 nm.

Figure S2 also provides an estimate of the effective saturation voltage for our LC-metasurface devices. In all subplots, the rate of resonance shift peaks around 4 $V_{p-p}$ and begins to plateau around 10 $V_{p-p}$. In the main text, we used a saturation voltage of 20 $V_{p-p}$ (10 $V_{rms}$) to provide an upper limit for our measured transmittance spectrum. While we observed a plateau around 10 $V_{p-p}$, we increased to 20 $V_{p-p}$ to be sure that we are completely in a region where the mode tuning as a function of voltage has saturated.

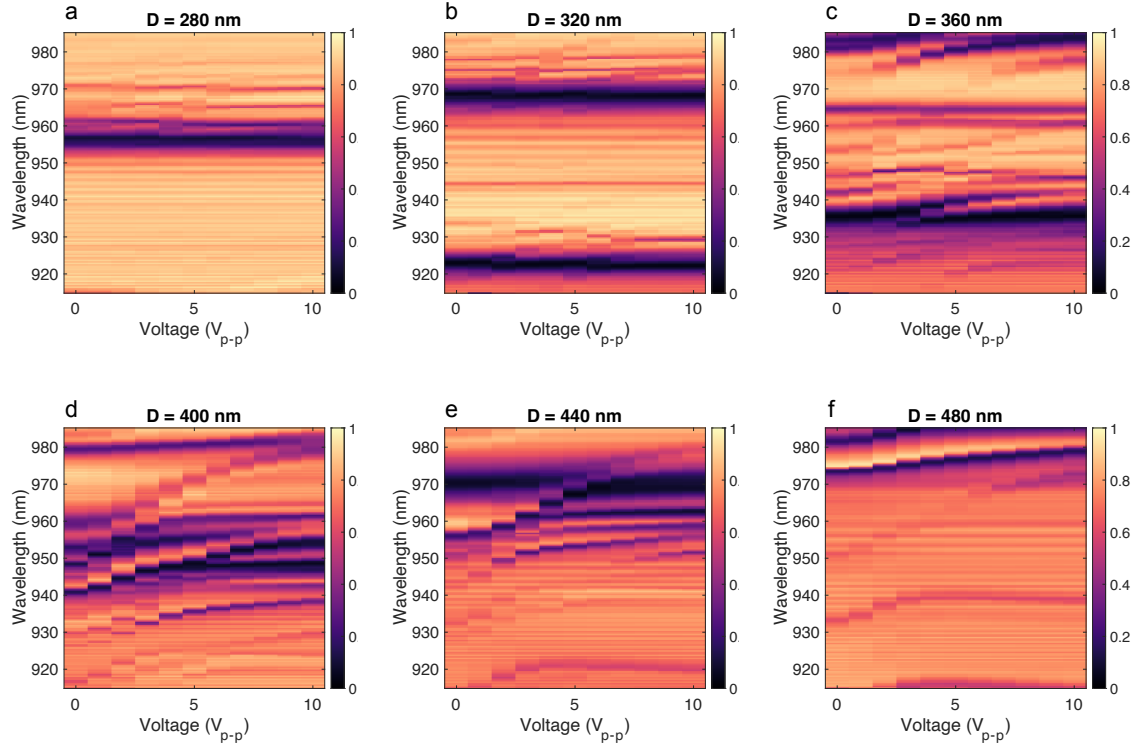

**Figure S2. Measured transmittance modulation of LC-infiltrated $TiO_2$ metasurfaces as a function of voltage. (a)-(f)** Results for metasurfaces with cylindrical nanopillars of $TiO_2$ with diameters, D, from 280 – 480 nm, stepping by 40 nm. Full device dimensions are outlined in Fig. 2 in the main text.

**Landau-de Gennes liquid crystal simulations**

The LC simulations performed in this work use the open-source python code, open-Qmin, developed as a collaboration between the research groups of Daniel Sussman (Emory University) and Daniel Beller (Johns Hopkins University) [1]. This code calculates a nematic director field, $\hat{n}(\boldsymbol{r})$, which describes the LC orientation in each point in space. As LCs have "head-tail" symmetry, meaning they feel the same force in a parallel vs antiparallel orientation, the director field is symmetric such that $\hat{n}(\boldsymbol{r}) = -\hat{n}(\boldsymbol{r})$. Because of this symmetry, it is numerically more convenient to describe the LC orientation using a symmetric second-rank tensor (3 x 3 in 3 dimensions). This tensor is the nematic orientation tensor, known as the Q-tensor, $\boldsymbol{Q}$. Open-Qmin solves for this Q-tensor at each lattice site in 3 dimensions to exactly specify the LC director field.

To obtain the final LC director field, open-Qmin numerically minimizes the Landau-de Gennes free energy, a functional of the Q-tensor:

$$F[\boldsymbol{Q}] = \int (f_{bulk} + f_{distortion} + f_{external})dV + \sum_{\alpha} \int (f_{boundary}^{\alpha})ds_{\alpha} \qquad (1)$$

For further details, readers are referred to the open-Qmin documentation [1]. The rest of this section will be dedicated to the simulation setup and boundary conditions used in this work.

For all simulations, we assumed a unit cell periodic in the x- and y-axes and fixed in the z-axis. The simulation region had dimensions of 576 nm x 576 nm x 2002.5 nm in the x, y, z dimensions, respectively. (Note: the lattice spacing in open-Qmin was fixed to be 4.5 nm. As a result, our unit cell dimensions vary slightly from those experimentally fabricated.) After defining the boundary conditions of our unit cell, we added a cylindrical surface of varying diameter (ranging from 280 – 480 nm) to represent our $TiO_2$ pillars.

Next, we defined the type of free energy anchoring conditions on each surface in our unit cell. Two anchoring conditions are possible in open-Qmin: oriented and degenerate planar. For oriented anchoring, LCs will prefer to align along some specified direction. For degenerate planar, LCs will prefer to align in the plane of the surface. In the degenerate planar configuration, the LCs can rotate freely as long as they are orthogonal to the surface normal. In our cell, we defined the top z = 2002.5 nm surface to have an oriented anchoring condition along the y-axis. This boundary condition approximates the rubbed alignment layer along the y-axis in our fabricated device. For all other surfaces ($TiO_2$ pillar and bottom substrate), we specified oriented anchoring condition along the surface normal. This condition was selected because of the HMDS surface treatment applied to the $TiO_2$ pillar and underlying $SiO_2$, which has been previously reported to exhibit a homeotropic (oriented to the surface normal) anchoring condition with E7 LCs [2].

After specifying the type of anchoring condition, we selected the strength of anchoring. While anchoring strengths can be measured experimentally, they are difficult to measure for 3D structures such as our $TiO_2$ nanopillars. As an estimate (taken from literature), we assumed an anchoring strength of $2 \times 10^{-4}$ J/m$^2$ for our rubbed polyimide alignment layer and $1 \times 10^{-4}$ J/m$^2$ for our HMDS coated $TiO_2$ pillars and bottom $SiO_2$ substrate [3]. As the exact value is unknown, we varied the anchoring strength of the HMDS coated $TiO_2$ pillars and bottom $SiO_2$ substrate to understand the sensitivity of this parameter and selected a moderate value which gave a good match with experiment.

Using these boundary conditions and material parameters, we then solved the Landau-de Gennes equations using open-Qmin and obtained the final Q-tensor which was then converted to a director field. From here, we were then able calculate the spatially varying DC permittivity of our cell or the spatially varying optical refractive index and import these fields into either electrostatic or optical frequency simulations.

**References**

1. Sussman, D. M. & Beller, D. A. Fast, Scalable, and Interactive Software for Landau-de Gennes Numerical Modeling of Nematic Topological Defects. Front Phys **7**, (2019).
2. Manyuhina, O. V., Tordini, G., Bras, W., Maan, J. C. & Christianen, P. C. M. Doubly periodic instability pattern in a smectic-A liquid crystal. Phys Rev E **87**, 050501 (2013).
3. De Cort, W., Beeckman, J., Claes, T., Neyts, K. & Baets, R. Wide tuning of silicon-on-insulator ring resonators with a liquid crystal cladding. Opt Lett **36**, 3876 (2011).